%% file: ms.tex
\documentclass[11pt]{article}
\pdfoutput=1

\usepackage[utf8]{inputenc}
\usepackage[T1]{fontenc}
\usepackage[USenglish]{babel}

\usepackage{jheppub}
\usepackage{amsfonts}
\usepackage{amsmath}
\usepackage{amssymb}
\usepackage{amsthm}

\usepackage{float}
\usepackage{placeins}
\usepackage{booktabs}

\usepackage{caption}
\usepackage{subcaption}
\usepackage{graphicx}

\usepackage{color}
\usepackage{siunitx}

\newcommand{\be}{\begin{equation}}
\newcommand{\ee}{\end{equation}}
\newcommand{\bea}{\begin{eqnarray}}
\newcommand{\eea}{\end{eqnarray}}

\newcommand{\Char}{\mathrm{Char}}
\newcommand{\Res}{\mathrm{Res}}
\newcommand{\SU}{\mathrm{SU}}
\newcommand{\Z}{\mathbb{Z}}
\newcommand{\R}{\mathbb{R}}

\newcommand{\mmmatrix}[9]{ \left(\! \begin{array}{ccc}#1 & #2 & #3\\ #4 & #5 & #6\\ #7 & #8 & #9\\ \end{array}\!\right) }
\newcommand{\unitmat}{1\!\!1}

\def\lsim{\mathrel{\rlap{\lower4pt\hbox{\hskip1pt$\sim$}}
    \raise1pt\hbox{$<$}}}         %less than or approx. symbol
\def\gsim{\mathrel{\rlap{\lower4pt\hbox{\hskip1pt$\sim$}}
    \raise1pt\hbox{$>$}}}         %greater than or approx. symbol

\title{Algorithmic Boundedness-From-Below Conditions for Generic Scalar Potentials}

\author[a]{Igor P. Ivanov,}
\author[b]{Marcel K\"opke,}
\author[b]{Margarete M\"uhlleitner}

\affiliation[a]{CFTP, Instituto Superior T\'ecnico, Universidade de Lisboa, av. Rovisco Pais 1, 1049-001 Lisboa, Portugal}
\affiliation[b]{Institute for Theoretical Physics, Karlsruhe Institute of Technology, Wolfgang-Gaede-Straße 1, 76131 Karlsruhe, Germany}

\emailAdd{igor.ivanov@tecnico.ulisboa.pt}
\emailAdd{marcel.koepke@gmail.com}
\emailAdd{margarete.muehlleitner@kit.edu}

\abstract{%
\input{abstract.tex}
}

\begin{document}

\maketitle

% sections in folder "sec"
\section{Introduction}
\label{sec:introduction}
\input{sec/introduction}

\section{Algorithmic Path to BFB Conditions}
\label{sec:idea}
\input{sec/idea}

\section{Elements of the Spectral Theory of Tensors}
\label{sec:spectral_theory}
\input{sec/spectral_theory}

\section{Applications}
\label{sec:applications}
\input{sec/applications}

\section{Discussion}
\label{sec:discussion}
\input{sec/discussion}

\bigskip

% acknowledgments in file "acknowledgments.tex"
{\bf Acknowledgments.}
\input{acknowledgments.tex}

%%%%%%%%%%%%%%%%%%%%%%%%%%%%%%%%%%%%%
%%%%%%%%%%%% Appendices %%%%%%%%%%%%%
%%%%%%%%%%%%%%%%%%%%%%%%%%%%%%%%%%%%%

\appendix

% appendix in folder "app"
\section{Polynomial Rings, Division and Resultants}
\label{app:resultants}
\input{app/resultants}

\bibliography{ms}
\bibliographystyle{JHEP}

\end{document}

%% file: abstract.tex
Checking that a scalar potential is bounded from below (BFB) is an ubiquitous
and notoriously difficult task in many models with extended scalar sectors.
Exact analytic BFB conditions are known only in simple cases.
In this work, we present a novel approach to algorithmically establish the BFB conditions for any polynomial scalar potential.
The method relies on elements of multivariate algebra, in particular, on resultants and on the spectral theory of tensors, which is being developed by the mathematical community.
We give first a pedagogical introduction to this approach, illustrate it with elementary examples, and then present the working Mathematica implementation publicly available at GitHub.
Due to the rapidly increasing complexity of the problem, we have not yet produced ready-to-use analytical BFB conditions for new multi-scalar cases.
But we are confident that the present implementation can be dramatically improved and may eventually lead to such results.

%% file: sec/introduction.tex
\subsection{The Problem}

Dealing with scalar potentials is one of the ubiquitous tasks one faces when building models beyond the Standard Model (SM).
Since the discovery of the Higgs boson in 2012~\cite{Aad:2012tfa,Chatrchyan:2012xdj}, we know that the Higgs mechanism, in some form, is at work.
What we do not know is whether it is as minimal as in the SM or if the SM-like $\SI{125}{\GeV}$ Higgs boson is the tip of the iceberg of a sophisticated scalar sector \cite{Ivanov:2017dad}.

When working with multiple interacting scalar fields, one usually builds a scalar potential and then finds its minimum to determine the vacuum expectation value configuration.
Before minimizing the potential, it has to be made sure that a global minimum exists in the first place.
Thus, one must verify that the potential is bounded from below (BFB).\footnote{%
% long footnote
To be precise, boundedness from below is a necessary but not sufficient condition for a minimum to exist.
Consider, for example, the following function of two real variables $x$ and $y$:
$$
V(x,y) = (xy-1)^2 + y^4 \text{ .}
$$
It is clearly bounded from below, as both terms are strictly non-negative, but it does not possess a global minimum.
As one moves along the hyperbole $xy = 1$ to large $x$ values, $V\to 0$ but never reaches zero.
However, we know of no multi-scalar example which makes use of this mathematical peculiarity.
Therefore, in this paper, the BFB conditions will be understood as equivalent to the existence of a minimum.
% end long footnote
}

At tree level, the scalar potential is written as a polynomial in scalar fields.
If one keeps the scalar interactions renormalizable, the polynomial degree of the potential is four.
By denoting the real scalar fields generically as $\phi_i$, $i = 1, \dots, n$, one can represent such a scalar potential as
\begin{equation}
\label{equ:generic_potential}
    V(\phi_i) = V_0 + Q_{i j k l} \phi_i \phi_j \phi_k \phi_l \text{ ,}
\end{equation}
where $V_0$ includes all lower-degree monomials and a summation over repeated indices is assumed.
At large quasiclassical values of the scalar fields, the quartic term dominates over the lower-degree terms.
Therefore, the condition for the potential $V$ to be bounded from below in the strong sense is equivalent to the requirement that 
\be
Q_{ijkl}\phi_i \phi_j \phi_k \phi_l > 0\quad \mbox{for all non-zero vectors $(\phi_i) \in \R^n$.}\label{BFB-condition}
\ee
Since the scalar potential depends on several free parameters, which we collectively denote $\{\Lambda_a\}$,
the BFB condition (\ref{BFB-condition}) carves out a region in the $\{\Lambda_a\}$-space.
If one wishes to build a model based on the potential, one must make sure the selected parameters correspond to a point
inside it. Thus, the task is to efficiently describe this region, preferably in terms of inequalities on the parameters $\{\Lambda_a\}$.

It is this task, in the general setting, that we want to attack in this work in an algorithmic fashion.

Before we move on, let us make a few clarifying comments.
First, a potential can be bounded from below even if there exist some flat directions of the quartic potential,
that is, subspaces of $\R^n$ in which the quartic term in (\ref{BFB-condition}) is exactly zero.
In this case, one needs to require that, within these subspaces, 
the lower-degree terms in the scalar potential grow and not decrease at large values of the fields. 
This situation was called in \cite{Maniatis:2006fs} stability in the weak sense.
Geometrically, it corresponds to the boundary of the BFB region in the $\{\Lambda_a\}$-space, which we have just described.
The solution to the BFB problem in the strong sense, Eq.~(\ref{BFB-condition}), 
is a prerequisite to establishing stability in the weak sense. 
Therefore, from now on, we focus only on the BFB problem in the strong sense.

Second, one can distinguish {\em necessary} BFB conditions and {\em sufficient} BFB conditions.
Necessary BFB conditions are the ones, which are truly unavoidable: their violation immediately drives 
the potential to be unbounded from below. However, satisfying a set of necessary conditions does not automatically
imply that the potential is BFB: the necessary conditions may be too weak for that.
Conversely, sufficient BFB conditions are safe: if a parameter set satisfies them, the potential is guaranteed to be BFB.
However, they may be overly restrictive: not satisfying a set of sufficient conditions does not automatically rule out 
a given parameter set. So, although a set of sufficient BFB conditions may be easy to establish and implement in numerical scans,
it will miss potentially interesting parts of the available parameter space.

What we are looking for is a set of BFB conditions which are, simultaneously, necessary and sufficient.
They are more difficult to establish than just a set of necessary and a different set of sufficient conditions,
but they incorporate the full information on the allowed parameter space in a given class of models.

Third, in quantum fields theory, quantum corrections can destabilize a potential that would be stable in the classical approximation.
Finding the quantum corrections to the classical potential and checking their effect on stability
is a separate issue which we do not address in this work.
Fortunately, in many cases, the main effect of quantum corrections can be absorbed
into running parameters of the potential $\{\Lambda_a\}$ without changing the polynomial structure of the potential.
In these cases, the mathematical task of establishing the BFB conditions remains unchanged.

\subsection{Overview of the Approaches to BFB Conditions}

Establishing the necessary and sufficient BFB conditions is a technical, but notoriously difficult problem
in any sophisticated multi-scalar theory.
There is no general, ready-to-use solution to this problem, and various
approaches have been proposed for particular scalar sectors.
Although our work does not rely on them, we find it instructive to give a brief overview
of these approaches.
We will explicitly give the potentials and denote their coefficients by $\Lambda_a$ instead of 
the more traditional notation $\lambda_a$ because $\lambda$'s will be reserved for the eigenvalues in the remainder of the text. 

In models with few degrees of freedom or few interaction terms, 
the exact BFB conditions can be established with straightforward algebra.
The convenient approach is to split the degrees of freedom in the scalar field space into ``radial'' and ``angular'' ones,
factor out the radial dependence of the quartic potential and explore the full domain of the angular coordinates.
For instance, if the quartic scalar potential depends on two fields $\phi_1$ and $\phi_2$, irrespective of their gauge quantum numbers, 
via the portal-type coupling
\be
V = \Lambda_1|\phi_1|^4 + \Lambda_2 |\phi_2|^4 + \Lambda_3 |\phi_1|^2|\phi_2|^2\,,
\ee
then one can parametrize $|\phi_1|^2 = r\cos\theta$, $|\phi_2|^2 = r\sin\theta$, 
with $0 \le \theta \le \pi/2$, and rewrite the potential as
\be
V = r^2(\Lambda_1 \cos^2\theta + \Lambda_2\sin^2\theta + \Lambda_3\sin\theta\cos\theta)\,,
\ee
which must be positive definite for all values of $\theta$. Since the angular dependence
can be written via the sine and cosine of the single angle $2\theta$,  
this requirement immediately leads to $\Lambda_1 > 0$,
$\Lambda_2 > 0$, and $\Lambda_3 + 2\sqrt{\Lambda_1 \Lambda_2} > 0$.

This approach was used, for example, back in 1978 \citep{IDMBFBConditions} to establish the BFB conditions 
for the two-Higgs-doublet model (2HDM) with unbroken $\Z_2$ symmetry,
which was later dubbed the Inert Doublet Model (IDM). 
This model uses two electroweak Higgs doublets $\phi_1$ and $\phi_2$, and its  
quartic potential has five terms: 
\be
    V = \Lambda_1 | \phi_1 |^4 + \Lambda_2 | \phi_2 |^4 
    + \Lambda_3 | \phi_1 |^2 | \phi_2 |^2 + \Lambda_4 | \phi_1^\dagger \phi_2 |^2 
    + \frac{\Lambda_5}{2} \left[ (\phi_1^\dagger \phi_2)^2 + (\phi_1^\dagger \phi_2)^2 \right]\,,\label{IDM-potential}
\ee
with all parameters being real. The BFB conditions are
\begin{equation}
\label{IDM-BFB}
    \Lambda_1 > 0\,, \quad \Lambda_2 > 0\,, \quad \Lambda_3 + 2 \sqrt{\Lambda_1 \Lambda_2} > 0\,,
    \quad \Lambda_3 + \Lambda_4 - | \Lambda_5 | + 2 \sqrt{\Lambda_1 \Lambda_2} > 0\,.
\end{equation}
In the most general 2HDM, which includes such interaction terms as $(\phi_1^\dagger\phi_2)(\phi_1^\dagger\phi_1)$,
this method runs into the difficulty of dealing with several competing angular functions of different periods.
It was only after the 2HDM potential was rewritten in the space of gauge-invariant
bilinears \cite{Nagel:2004sw,Ivanov:2005hg,Maniatis:2006fs,Nishi:2006tg},
that the BFB conditions could be established. 
They were first presented in the form of an algebraic algorithm \cite{Maniatis:2006fs}
and later written in compact closed form \cite{Ivanov:2006yq}
as inequalities imposed not on the parameters $\Lambda_a$ themselves but on four eigenvalues $\hat{\Lambda}_i$
of a real symmetric $4\times 4$ matrix $\Lambda_{ij}$ which encodes all quartic interaction terms.
The form of these conditions is very simple and basis-invariant,
\be
\hat{\Lambda}_0 > 0\,, \quad \hat{\Lambda}_0 > \hat{\Lambda}_{1,2,3}\,, \label{2HDM-BFB}
\ee
but checking them within a specific 2HDM requires first finding these eigenvalues,
though this step can be easily implemented in numerical scans of the parameter space.

A somewhat similar systematic method of deriving the exact BFB conditions exists for models, 
in which the Higgs potential can be written in terms of independent {\em positive-definite} field bilinears $r_i$.
In this case, the quartic potential can again be rewritten as a quadratic form
$V = \Lambda_{ij} r_i r_j$, but its positive definiteness must be insured only in the first orthant $r_i \ge 0$.
These conditions are known as copositivity (conditional positivity) criteria. 
They were developed in \cite{Copositivity,Copositivity2,Kannike:2016fmd} and applied to such cases as 
some 2HDMs, singlet-doublet models, models with $\Z_3$ symmetric scalar dark matter, and left-right symmetric models.

Beyond two Higgs doublets, in the general $N$-Higgs-doublet model (NHDM), 
the exact BFB conditions in closed form are still not known. Several attempts to attack the problem 
with the bilinear space formalism \cite{Maniatis:2006fs,Nishi:2007nh,Ivanov:2010wz,Maniatis:2014oza,Maniatis:2015gma}
did not culminate in a closed set of inequalities.
The technical challenge is that, with $N$ Higgs doublets, the space of bilinears $r_a$, $a = 1, \dots, N^2-1$,
does not span the entire $\R^{N^2-1}$ space but only a lower-dimensional algebraic manifold, 
which is described with a series of polynomial constraints. Positive-definiteness of a quadratic form 
on a complicated algebraic manifold cannot be decided with linear algebra
and requires algebraic-geometric tools, that have not been found yet.

For larger gauge symmetries and for scalars in higher-dimensional representations, 
it is appropriate to analyze the scalar potential not in the scalar fields space
but in the space of gauge orbits.
This approach flourished in 1980's
with the advent of Grand Unification models, see, for example, \cite{Kim:1981xu,Kim:1981jj,Abud:1981tf,Abud:1983id}
and a short historical overview in \cite{Kannike:2016fmd}.

In specific multi-Higgs models, in which large continuous or discrete symmetry groups
dramatically simplify the potential, the exact conditions can be established
\cite{Branco:1983tn,Klimenko1985,Toorop:2010ex,Degee:2012sk,Maniatis:2015kma,Heikinheimo:2017nth}. 
We mention, in particular, the method developed in \cite{Degee:2012sk,Heikinheimo:2017nth}
to rewrite the Higgs potential as a {\em linear} combination of new variables, the group-invariant quartic field combinations,
and to determine the exact shape of the space spanned by these variables.
This method is similar to the so-called linear programing, and 
it gives the BFB constraints directly from the description of the shape of the space available.

In certain cases, when the exact necessary and sufficient conditions are not known
but a parameter scan still needs to be performed, it may be enough to write down a set of 
{\em sufficient} conditions. They may be overly restrictive, but if a point satisfies them,
the potential is guaranteed to be positive-definite.
An example of such conditions was given for a specific 3HDM in \cite{Ferreira:2017tvy}.
The idea is to pick up all terms with ``angular'' dependence in the scalar field space
and find a lower bound for each term separately.
For example, if the potential contains a term $(\phi_1^\dagger \phi_2)(\phi_1^\dagger \phi_3)$
with real coefficient $\Lambda$, one can place the following lower boundary on it
in the $r_i \equiv |\phi_i|^2$ space:
\be
\Lambda (\phi_1^\dagger \phi_2)(\phi_1^\dagger \phi_3) \ge 
-|\Lambda| r_1 \sqrt{r_2r_3} \ge - |\Lambda|r_1(r_2+r_3)/2\,.
\ee
In this way, the original potential $V$ can be limited from below by another potential $\tilde V$,
which is a quadratic form in terms of $r_i$ and for which the copositivity criteria are applicable.

\bigskip

In this work we present an algorithm, which in principle solves the problem in a generic setting.
The algorithm uses elements of the theory of resultants and of the recently developed spectral theory of tensors.
However, solving the problem in principle is quite different from solving it in practice.
To our best knowledge, the approach was only briefly mentioned in \cite{Kannike:2016fmd} 
but was not developed any further nor implemented in any code.
We have implemented the method in a computer-algebra code, which is available at GitHub \cite{BFB}, 
and tested it in cases, in which analytical solutions already exist.
The complexity of the algorithm implementation grows so fast that, with limited computer resources, 
we could not apply it to cases where the results are not yet known.

This does not imply, of course, that this direction is a dead-end.
The method itself is innovative but the specific algorithm we propose is clearly not optimal. 
We believe that with additional efforts, it can be seriously improved
and may eventually produce a ready-to-use solution in various popular classes of multi-scalar models,
such as the general 3HDM.

The structure of the paper is the following.
In the next Section, we present our strategy and formulate the algorithm.
Section~\ref{sec:spectral_theory} contains an introduction to the spectral theory of tensors,
its application to the BFB problem, and describes a practical algorithm to calculate the characteristic polynomial
of a symmetric tensor.
In Section~\ref{sec:applications}, we show how this method works.
We first do it with two elementary examples, in which all calculations can be performed manually,
and then apply the computer-algebra package to the case of a $\Z_2$-symmetric 2HDM, where the BFB conditions are known.
We find agreement of the results, which serves as a check of the validity of our algorithm.
We end with a discussion in Section~\ref{sec:discussion} of how the algorithm can be improved in the future and draw conclusions.
The appendix contains a pedagogical introduction to polynomial rings and polynomial division with an application to the theory of resultants.

%% file: sec/idea.tex
The BFB condition (\ref{BFB-condition}) is formulated in terms of positive-definiteness of the real fully symmetric 
order-four tensor $Q_{i j k l}$ in the entire space of real non-zero vectors $\phi_i$, $i = 1, \dots, n$.
If the order of the tensor were not four but two, $M_{ij}$, then its positive definiteness 
in the entire space of non-zero vectors $\phi_i$ could be easily established
with elementary linear algebra. 
One first views the tensor $M_{i j}$ as a linear operator acting in the space $\R^n$ of vectors $\phi_i$,
and asks for its eigenvalues and eigenvectors:
\be
M_{ij}\phi_j = \lambda \cdot \phi_i\,.
\ee 
For a real symmetric $M_{ij}$, there are $n$ real eigenvalues, which can be found from the characteristic equation 
\be
\Char(M,\lambda)=\det(M - \lambda \cdot \unitmat) = 0\,.
\ee
The tensor $M_{ij}$ is positive definite if and only if all $\lambda_i$ are positive: $\lambda_i > 0$. 
The calculation of the determinant is done through well-known algorithms, 
and it produces a polynomial for $\lambda$ of degree $n$, whose coefficients
are multi-linear functions of each individual entry of the matrix $M_{ij}$.

The critical complication of recasting the BFB condition (\ref{BFB-condition})
into constraints on the order-four tensor $Q_{ijkl}$ lies precisely in the fact that it has higher order.
Linear algebra is of no use anymore. One needs to develop a theory that generalizes the above 
chain ``characteristic equation $\to$ determinant $\to$ eigenvalues $\to$ positivity'' to the case of higher-order tensors,
and to supplement the general theory with efficient algorithms.

This theory exists and is known as the {\em spectral theory of tensors}.
Although the issue must have been discussed earlier,
it was only in 2005 that Lim \citep{Lim2005} and Qi \citep{Qi2005}, independently from each other,
constructed fruitful generalizations of spectral theory to higher-order tensors.
These and subsequent works gave a huge boost to the field, resulting in 
numerous applications in various branches of pure and applied mathematics, 
for a brief review and a pedagogical introduction, see \cite{QiReview} and the very recent book \citep{QiBook}.
We will also provide an introduction in the following section.
For the moment, we outline the general strategy.

There indeed exists a way --- in fact, several ways --- of generalizing eigenvalues and eigenvectors
to tensors $Q_{i_1 i_2 \cdots i_m}$ of order $m$ (in our case, $m=4$). 
They can be written as a system not of linear but of polynomial equations of degree $m-1$.
The eigenvalues $\lambda$ are again determined by a characteristic equation $\Char(Q,\lambda) = 0$.
However it is calculated not via the determinant but via the {\em resultant} of a system of equations.
The resultant is a polynomial in $\lambda$, whose coefficients are polynomial --- and not just linear --- functions of the entries of the tensor $Q$.
It is much more complicated than the determinant; in particular, its degree can be much larger than $n$.
However, there are algorithms for calculating resultants, that can be implemented in computer-algebra codes.

Once the resultant is found, its roots give all the eigenvalues $\lambda$.
It may happen that some of these real eigenvalues may correspond to complex eigenvectors only.
It just so happens that such eigenvalues can be disregarded with respect to positive definiteness.
Hence, we focus only on those eigenvalues that produce real eigenvectors.
The tensor $Q$ is positive definite if and only if all of these remaining eigenvalues are positive.

From a computational point of view, the most challenging and computer-time consuming step is calculating the resultant
for a given model.
Just like for determinants, there exists a recursive algorithm, 
but for non-linear equations its complexity grows dramatically with the number of equations, variables and the degree of the polynomials.
The coefficient of the characteristic polynomial may easily become so large that usual computer packages are incapable 
of manipulating such coefficients.
Specialized algebraic-geometric packages are needed for this purpose. 

Once the resultant is found in its analytic form, it can be used for any set of parameters.
Checking the positivity of those of its roots that correspond to real eigenvectors can be done numerically
in short time.
In this way, even if the BFB conditions cannot be written in a nice closed form,
they can easily be implemented in numerical scans of the parameter space.

%% file: sec/spectral_theory.tex
In this Section, we introduce the basics of the spectral theory of tensors, which will be needed 
to describe the algorithm we implemented.
The presentation is based on the theory developed by Qi \citep{Qi2005}.
A much more detailed introduction can be found in the review \citep{QiReview} and the book \citep{QiBook}.

\subsection{Eigenvalues and Positive Definiteness} \label{sec:bounded:eigenvalues}

Let $Q$ be a real, fully symmetric tensor of order $m$ over the vector space $\mathbb{C}^n$.
The elements of this vector space are denoted by $\vec x$.
Although we will eventually be interested in this tensor over the real vector space $\R^n$,
we need the complex space for the intermediate steps.

We call $\lambda \in \mathbb{C}$ an eigenvalue of $Q$ if the system of equations
\be
\label{equ:bounded:eigenvalue}
    Q_{i_1 i_2 \dots i_m} \cdot x_{i_2} \dots x_{i_m} = \lambda \cdot x_{i_1}^{m -1}
\ee
has non-trivial solutions $\vec{x} \in \mathbb{C}^n \setminus \{0\}$.
These solutions $\vec{x}$ are then called eigenvectors.
Notice that in Eq.~(\ref{equ:bounded:eigenvalue}) all indices apart from $i_1$ are summed over. 
The index $i_1 = 1, \dots, n$ is an open index; it labels the $i_1$-th equation. 
Thus, Eq.~(\ref{equ:bounded:eigenvalue}) represents a system of $n$ homogeneous polynomial equations 
of degree $m-1$ in $n$ variables $x_i$.
The total number of eigenvalues, including multiplicity, is \cite{Qi2005}
\be
N_{Q} = n(m-1)^{n-1}\,.\label{counting-eigenvalues}
\ee
For $m = 2$, the definition of Eq.~\eqref{equ:bounded:eigenvalue} reduces to the eigensystem of square matrices,
and the total number of eigenvalues is equal to $n$.

Even if the tensor $Q$ is real and symmetric, its eigenvalues and eigenvectors can be complex.
A {\em real} eigenvector is called an H-eigenvector; its associated eigenvalue --- which is unavoidably real --- is called an H-eigenvalue. 
If one restricts the vector space from $\mathbb{C}^n$ to $\R^n$, then only the H-eigenvalues and H-eigenvectors survive.
A key theorem that links the spectral theory of tensors with the BFB conditions is due to Qi \cite{Qi2005}. 
Suppose the real symmetric tensor $Q$ is of even order: $m = 2 k$.
Then H-eigenvalues exist, and $Q$ is positive definite, 
\be
    Q_{i_1 i_2 \dots i_m} \cdot x_{i_1} x_{i_2} \dots x_{i_m} > 0 \quad \mbox{for all} \ \vec x \in \R^n \setminus \{0\}\,, \label{positivity-general}
\ee
if and only if all of its H-eigenvalues are positive.
The task of establishing the BFB conditions reduces to finding the H-eigenvalues of the tensor $Q$.

We remark here that, in contrast to the eigenvalues of matrices, the eigenvalues of tensors defined according to (\ref{equ:bounded:eigenvalue})
are not invariant under general basis rotations. 
In particular, the H-eigenvalues are not invariant under generic $O(n)$ rotations. 
It turns out, however, that the property of 
all H-eigenvalues being positive {\em is} $O(n)$-invariant. 
It is this property that makes them a useful indicator of positive-definiteness in any basis.

We note in passing that in certain problems, where the $O(n)$ invariance of eigenvalues is crucial, 
one can adopt another definition of eigenvalues, which is manifestly basis-change invariant \citep{Lim2005}. 
The problem of positive definiteness of the tensor $Q$ can also be formulated in terms of positivity of these new eigenvalues.
In this work, we prefer to stick to the H-eigenvalues, as their application seems to be more straightforward.

\subsection{Characteristic Polynomial and Resultant} \label{sec:bounded:charpoly}

In order to find eigenvalues of the tensor $Q$, let us rewrite the system of coupled homogeneous polynomial
equations (\ref{equ:bounded:eigenvalue}) in the following form:
\begin{equation}
\begin{split}
    f_1 & = Q_{1 i_2 \dots i_m} \cdot x_{i_2} \dots x_{i_m} - \lambda \cdot x_1^{m-1} = 0 \\
    f_2 & = Q_{2 i_2 \dots i_m} \cdot x_{i_2} \dots x_{i_m} - \lambda \cdot x_2^{m-1} = 0 \\
    & \qquad \qquad \vdots \\
    f_n & = Q_{n i_2 \dots i_m} \cdot x_{i_2} \dots x_{i_m} - \lambda \cdot x_n^{m-1} = 0\,. \label{f1f2fn}
\end{split}
\end{equation}
In that way, we simply ask for non-trivial ($\vec x \not = 0$) solutions to $n$ coupled, homogeneous polynomial equations in $n$ variables.
In particular, we want to know for which values of $\lambda$ such solutions exist.
For any system of homogeneous polynomials $f_1, \dots, f_n$ of $n$ variables $x_1, \dots, x_n$, 
there always exists a polynomial in the coefficients of $f_1, \dots, f_n$, called the resultant $\Res(f_1, f_2, \dots, f_n)$, 
with the following property \cite{Qi2005}: non-zero solutions to $f_1 = 0, \dots, f_n = 0$ exist if and only if $\Res(f_1, f_2, \dots, f_n) = 0$.
In the case of Eqs. (\ref{f1f2fn}), the coefficients of $f_i$ contain $\lambda$.
The resultant $\Res(f_1, f_2, \dots, f_n)$ can then be viewed as a single polynomial in $\lambda$ whose coefficients depend on the entries of the tensor $Q$.
It is called the characteristic polynomial $\Char(Q,\lambda)$, and its roots give all the eigenvalues of the tensor $Q$.
Just as for determinants, the value of $\Char(Q,\lambda = 0)$ is equal to the product of all eigenvalues.

Resultants are much more difficult to calculate than determinants.
In fact, for the fields $\mathbb{Q}$, $\mathbb{R}$ and $\mathbb{C}$ the calculation is at least NP-hard \citep{NPhardResultant}.
Every NP-problem has an algorithm for which the execution time scales exponentially with the input.
The calculation time is thus extremely sensitive to the number $n$ of polynomials and to their respective degrees.

Multivariate resultants were first studied by Macaulay \citep{Macaulay1902}.
Due to him there is an algorithm that expresses the resultant as a quotient of the determinants of two matrices.
The size of these two matrices grows rapidly with the number $n$ and the polynomial degrees $\deg (f_i)$,
which renders this algorithm not very space efficient.
A more economical algorithm can be found in \citep[theorem 3.4]{UsingAlgebraicGeometry}.
It uses a recursive approach that we present now. 
Readers wishing to refresh their knowledge about the ring of polynomials and polynomial division can consult the Appendix \ref{app:resultants}.

\subsection{An Explicit Resultant Algorithm}\label{sec:bounded:resultantalgo}

Given homogeneous polynomials $f_1, f_2, \dots, f_n \in \mathbb{C}[x_1, x_2, \dots, x_n]$ with degrees $d_i := \deg (f_i)$, 
we define two sets of new polynomials
\begin{equation}
\label{equ:bounded:newpoly}
\begin{split}
    \bar{f}_i & = f_i(0, x_2, \dots, x_n) \\
    F_i & = f_i(1, x_2, \dots, x_n)\,.
\end{split}
\end{equation}
The polynomials $\bar{f}_i$ are again homogeneous and of the same degrees $d_i$ but of $n-1$ variables.
One can use $n-1$ of them to define the smaller resultant $\Res(\bar{f}_2, \dots, \bar{f}_n)$.
If $\textrm{Res}(\bar{f}_2, \dots, \bar{f}_n) \neq 0$, one has
\begin{equation}
\label{equ:bounded:recursiveresultant}
    \textrm{Res}(f_1, f_2, \dots, f_n) = \left( ~ \textrm{Res}(\bar{f}_2, \dots, \bar{f}_n) ~ \right)^{d_1} \cdot \det M_1\,.
\end{equation}
Here, $d_1$ is the degree of the eliminated polynomial $f_1$,
and the matrix $M_1$ is defined by the map
\begin{equation}
\label{equ:bounded:resultantmap}
    M_1: [r] \mapsto [r] \cdot [F_1] = [r \cdot F_1]\,,
\end{equation}
with the quotient ring $\mathbb{C}[x_2, \dots, x_n] / \langle F_2, \dots, F_n \rangle$ viewed as a complex vector space of dimension $D = d_2 \times \cdots \times d_n$ with elements $[r]$ and $[F_1]$.

Let us explain the last statement in simple terms. 
It says that we need to consider the remainders $r$ which we get after dividing 
all possible polynomials in $x_2, \dots, x_n$ by the ring ideal constructed with the generating polynomials $F_2, \dots, F_n$.
These remainders form a vector space, and a basis for this vector space must be found.
The basis vectors (independent remainders) can be further multiplied by the polynomial $F_1$ --- the one dropped in the construction of the ideal --- and the results can be again reduced to the remainders and expanded in the same basis.
Thus, $F_1$ acts as a linear map in this space, and we describe it with the matrix $M_1$, 
whose determinant we calculate. 

In technical terms, 
we first build the monomial basis of this vector space by scanning through all possible monomials
$m_a = x_2^{e_2} \dots x_n^{e_n}$ of ascending total degree $\deg (m_a) = \sum_{i=2}^n e_i = 0, 1, 2,$ etc.
Then we divide all monomials by the ideal $\langle F_2, \dots, F_n \rangle$,
for which we first need to find the Gröbner basis $G_i$ (see the brief introduction in Appendix \ref{app:resultants}).
At the end, we obtain $D = d_2 \times \cdots \times d_n$ unique, non-zero, linear independent monomial 
remainders $r_a$ which serve as basis vectors $[r_a]$ of the quotient ring viewed as vector space.
The same division is repeated for the polynomials $r_a \cdot F_1$ whose remainders $[r_a \cdot F_1]$
can be expanded in this basis
\be
    [r_a \cdot F_1] = \sum_{b=1}^D ~ [r_b] \cdot (M_1)_{b a} ,.
\ee
In this way we obtain the desired square matrix $M_1$ and calculate its determinant.

One can recursively repeat the procedure $n-1$ times
to end up with the resultant of a single homogeneous polynomial $\tilde{f}_n$ in one variable $x_n$ of degree $d_n$.
The only possible form for this polynomial is
\begin{equation}
    \tilde{f}_n (x_n) = \alpha \cdot x_n^{d_n}\,,\label{final-resultant}
\end{equation}
with some $\alpha \in \mathbb{C}$.
By definition, the resultant $\textrm{Res}(\tilde{f}_n)$ is zero if and only if there are non-trivial solutions to $\tilde{f}_n = 0$. 
Therefore, 
\begin{equation}
    \textrm{Res}(\tilde{f}_n) = \alpha\,.
\end{equation}
Hence, after $n - 1$ steps the calculation of the resultant terminates with a trivial relation.

If it happens that one of $\bar{f}_i \equiv 0$, $i = 1, \dots, n$, 
then we must eliminate it instead of $\bar{f}_1$ and proceed further. 
But it may also happen that two or more among $\bar{f}_i \equiv 0$.
In this case, we get no more than $n-2$ polynomial conditions $\bar{f}_i=0$ on $n-1$ variables,
so that the system becomes underdetermined, and non-trivial solutions always exist,
which implies that $\textrm{Res}(\bar{f}_2, \dots, \bar{f}_n) = 0$.

In the following section and in the appendix, we give a few examples of how this algorithm works.

%% file: sec/applications.tex
\subsection{Elementary Example 1}

We start with the simplest possible example: the quadratic potential in two variables
\be
V(x_1,x_2) = a x_1^2 + 2b x_1x_2 + cx_2^2 \equiv Q_{ij}x_i x_j\,.\label{example1}
\ee
The eigenvalues are defined according to 
\bea
f_1 := && Q_{1j} x_j - \lambda x_1 = ax_1 + bx_2 - \lambda x_1 = 0\,,\nonumber\\
f_2 := && Q_{2j} x_j - \lambda x_2 = bx_1 + cx_2 - \lambda x_2 = 0\,.
\eea
These polynomials are of degrees $d_1 = d_2 = 1$.
According to the algorithm, we build two other polynomial sets:
\be
\bar{f}_1 := bx_2\,, ~~~~~~~~~~~~~~~~~ \bar{f}_2 := (c - \lambda) x_2\,, ~~~~~
\ee
and 
\be 
F_1 := a-\lambda+bx_2\,,\qquad F_2 := b + (c - \lambda) x_2\,,
\ee
and calculate the resultant as
\be
\Res(f_1,f_2) = \left( ~ \Res(\bar{f}_2) ~ \right)^{d_1} \cdot \det M_1\,. \label{example1-step1}
\ee
In the ring of all polynomials in $x_2$, we define the ideal $\langle F_2\rangle$,
and need to describe the space of remainders $r$ of polynomial division by the ideal $\langle F_2\rangle$.
This ideal is generated by the single polynomial, so there is no need to search for the Gröbner basis.
This space is one-dimensional, $D=d_2=1$, and the real unit $1$ can serve as the basis vector in this space.
The polynomial $F_1$ can be divided by this ideal giving the following remainder $r$:
\be
F_1 = a-\lambda+bx_2 = \frac{b}{c-\lambda} F_2 + \left( a-\lambda - \frac{b^2}{c-\lambda} \right) \equiv q F_2 + r\,.
\ee
Thus, the matrix $M_1$ is just a single number describing the linear map $[1] \to [1 \cdot F_1] = [r]$,
giving $M_1 = a-\lambda - b^2/(c-\lambda)$. 
Finally, according to (\ref{final-resultant}), the resultant $\Res(\bar{f}_2) = c-\lambda$.
Therefore, the total resultant in Eq.~(\ref{example1-step1}) is 
\be
\Res(f_1,f_2) = (c-\lambda)^{1}\cdot \left( a-\lambda - \frac{b^2}{c-\lambda} \right) = 
(c-\lambda)(a-\lambda) - b^2\,,
\ee
which coincides with the usual determinant of the matrix $(Q - \lambda \cdot \unitmat)$. 
By setting this resultant to zero, we obtain the characteristic equation, whose roots give the eigenvalues $\lambda$:
\be
\lambda_{1,2} = \frac{1}{2} \left( a+c \pm \sqrt{(a-c)^2 + 4 b^2} \right)\,.
\ee
These roots are real and correspond to real eigenvectors, therefore they qualify as H-eigenvalues.
The BFB conditions for the potential (\ref{example1}) are $\lambda_1 > 0$, $\lambda_2 > 0$.
One can recast these conditions into $\lambda_1 + \lambda_2 > 0$ and $\lambda_1 \lambda_2 > 0$,
which are then translated into the usual expressions $a > 0$, $c> 0$, and $ac - b^2 > 0$.

\subsection{Elementary Example 2}

The previous calculation was so simple because (1) we needed just one iteration, 
(2) the vector space of the remainders was one-dimensional,
(3) the polynomial equations were of degree 1. Let us now consider a slightly more elaborate example:
\be
V(x_1,x_2) = a x_1^4 + 2b x_1^2x_2^2 + cx_2^4 \equiv Q_{ijkl}x_i x_j x_k x_l\,.\label{example2}
\ee
The standard treatment of this potential resorts to the so-called copositivity criteria \cite{Copositivity}.
One defines new variables $z_1 = x_1^2$, $z_2=x_2^2$, and rewrites the potential as a quadratic form in 
terms of $z_1$ and $z_2$. Then one asks for the positive definiteness of this quadratic form not on the entire 
$(z_1,z_2)$ real plane but only in the first quadrant, $z_1, z_2 \ge 0$.
The final result is similar to the previous case with the third condition being more relaxed:
\be
a > 0, \quad c > 0, \quad \sqrt{ac} + b > 0\,,\label{copositivity-result}
\ee
which implies that $b$ can now be arbitrarily large provided it is positive.

Let us rederive these results via resultants.
The eigenvalues are defined according to 
\bea
f_1 := && ax_1^3 + bx_1x_2^2 - \lambda x_1^3 = 0\,,\nonumber\\
f_2 := && bx_1^2x_2 + cx_2^3 - \lambda x_2^3 = 0\,.\label{example2-system}
\eea
These polynomials are of degrees $d_1 = d_2 = 3$. The two auxiliary polynomial sets are
\be
\bar{f}_1 \equiv 0\,, ~~~~~~~~~~~~~~~~~~~~~ \bar{f}_2 := (c - \lambda) x_2^3\,, ~~~~~~~~
\ee
and 
\be 
F_1 := a-\lambda+bx_2^2\,,\qquad F_2 := b x_2 + (c - \lambda) x_2^3\,.
\ee
The ideal $\langle F_2 \rangle$ is again generated by a single polynomial in one variable,
and we do not need to search for the Gröbner basis.
The vector space of remainders of the polynomial division of all polynomials in $x_2$ 
by this ideal is three-dimensional.
The basis vectors can be chosen $r_1 = 1$, $r_2 = x_2$, $r_3 = x_2^2$.
Higher powers of $x_2$ can be divided giving remainders in this space;
for example
\be
x_2^3 = \frac{1}{c-\lambda} F_2 + \left( - \frac{b}{c-\lambda} \right) x_2\,,
\ee 
which is equivalent to $-b/(c-\lambda)\cdot r_2$.
We can then calculate the action of $F_1$ in this space:
\bea
1 \cdot F_1 &=& a-\lambda + b x_2^2 = (a-\lambda) \cdot r_1 + 0 \cdot r_2 + b \cdot r_3\,,\nonumber\\[1mm]
x_2 \cdot F_1 &=& (a-\lambda) x_2 + b x_2^3 = 0 \cdot r_1 + \left( a-\lambda - \frac{b^2}{c-\lambda} \right) \cdot r_2 + 0 \cdot r_3\,,\nonumber\\[1mm]
x_2^2 \cdot F_1 &=& (a-\lambda) x_2^2 + b x_2^4 = 0 \cdot r_1 + 0 \cdot r_2 + \left( a-\lambda - \frac{b^2}{c-\lambda} \right) \cdot r_3\,.
\eea
The matrix $M_1$ is
\be
M_1 = \mmmatrix{a-\lambda}{0}{0}{0}{q}{0}{b}{0}{q}\,,\quad \mbox{where}\quad q := a-\lambda - \frac{b^2}{c-\lambda} \,.
\ee
Knowing that $\Res(\bar{f}_2) = c-\lambda$, we can calculate the full resultant as
\bea
\Res(f_1,f_2) &=& \left( ~ \Res(\bar{f}_2) ~ \right)^{d_1}\cdot \det M_1 = (c-\lambda)^3 \cdot (a-\lambda) \left( a-\lambda - \frac{b^2}{c-\lambda} \right)^2
\nonumber\\
&=& (c-\lambda)(a-\lambda)\left[(c-\lambda)(a-\lambda) - b^2\right]^2\,. \label{example2-resultant}
\eea
Solving $\textrm{Char}(Q,\lambda) = 0$ yields six eigenvalues in accordance with Eq.~(\ref{counting-eigenvalues}):
\be
    \lambda_1 = a\,, \quad \lambda_2 = c\,, \quad
    \lambda_{3,4} = \lambda_{5,6} = \frac{1}{2} \left( a + c \pm \sqrt{(a-c)^2 + 4b^2}  \right)\,.\label{example2-eigenvalues}
\ee
all of which are always real.
In order to find which of them are relevant for the BFB check,
we need to find their eigenvectors. This can be done by substituting eigenvalues 
back into the original equations (\ref{example2-system}).
We find that $\lambda_1$ corresponds to $\vec x \propto (1,0)$,
$\lambda_2$ corresponds to $\vec x \propto (0,1)$. 
Thus, they qualify for H-eigenvalues and produce conditions $a> 0$ and $c> 0$.

For the remaining eigenvalues, the discussion requires some care.
If $b=0$, no additional eigenvalues appear; thus, we can safely consider $b \not = 0$.
In this case, the eigenvectors lie on the rays $x_2 = k \cdot x_1$
with the proportionality coefficient defined by
\be
k^2 = \frac{1}{2b} \left( c -a \pm \sqrt{(a-c)^2 + 4b^2}  \right)\,,
\ee
where the $\pm$ sign is the same as in (\ref{example2-eigenvalues}).
Since the square root is always larger than $|c-a|$, we always get one positive and one negative expressions
for $k^2$. Since we are looking for the real solutions, $k$ must be real,
and we always keep only one $k^2$, depending on the sign of $b$. Thus, we get the additional H-eigenvalue:
\bea
b>0 &\quad\Rightarrow\quad& \lambda = \frac{1}{2} \left[a + c + \sqrt{(\cdot)}\right]\,,\quad 
k^2 = \frac{1}{2b} \left[ \sqrt{(\cdot)} + (c-a) \right]\,,\nonumber\\
b<0 &\quad\Rightarrow\quad& \lambda = \frac{1}{2} \left[a + c - \sqrt{(\cdot)}\right]\,,\quad 
k^2 = \frac{1}{2|b|} \left[ \sqrt{(\cdot)} - (c-a) \right]\,,\label{example2-extra-H}
\eea
where $\sqrt{(\cdot)}$ denotes $\sqrt{(a-c)^2 + 4b^2}$.
This additional H-eigenvalue must also be positive for the potential to satisfy BFB conditions.
However, in the former case, $b>0$, the conditions we have already established $a >0$, $c>0$,
guarantee that this extra $\lambda$ is positive. No extra constraint is needed in this case.
In the latter case, $b < 0$, the condition $\lambda >0$ is a new one and it restricts the absolute value 
of the negative parameter $b$: $|b| < \sqrt{ac}$. In this way, we recover the copositivity result (\ref{copositivity-result}).

We can draw several observations from this example. 
First, we see that the degree of the characteristic polynomial quickly grows for non-linear equations.
Fortunately, we had to perform only one iteration in this example, and the degree stopped at six.
In more elaborate situations, even with two iterations, the degree will grow very fast.
At each iteration, the resultant is factorized into a secondary resultant and a determinant of a matrix $M$.
However it does not imply that the final expression for the resultant could be easily factorized into these blocks.
We saw that $\det M_1$ was {\em not} a polynomial in $\lambda$ on its own because it contained $\lambda$ in the denominator.
It required two extra powers of $c-\lambda$ to become a polynomial.
Therefore, for situations slightly more sophisticated than the elementary examples considered,
we may easily run into higher-order polynomials in $\lambda$ whose solutions cannot be written in closed algebraic form.

This leads us to the conclusion that one should abandon the hope to represent the BFB conditions
in such elaborate situations in terms of explicit inequalities placed on the parameters of the potential. 
The final analytical form of the exact BFB conditions will be $\Char(Q, \lambda) = 0$,
and one would need to resort to numerical methods to find all real solutions of the characteristic equation.
Fortunately, numerically solving polynomial equations in a single variable can be done in short time even for very high-degree polynomials.

Another observation is that eigenvalues themselves do not provide the final answer;
one also needs to check the corresponding eigenvectors.
Whether a given real eigenvalue is an H-eigenvalue or not depends on the numerical values of the tensor entries.
This is an additional complication for the fully analytic treatment of the problem but it can be resolved
in reasonable time with numerical methods.

We wrap up this example by noticing that even if one considers, instead of Eq.~\eqref{example2}, 
the most general quartic polynomial in two real variables, the resultant can still be found analytically
with the same strategy.
This case, however, has also been studied previously, \cite{Kannike:2016fmd}.

\subsection{Implementation}

The above two elementary examples were simple enough to be done by hand.
Although the calculations become much more involved in less trivial examples, 
the algorithm remains unchanged and can be implemented in a computer-algebra code.
We did it within the Mathematica \citep{Mathematica} and Macaulay2 \citep{M2} platforms, 
and our Mathematica package BFB \citep{BFB} is publicly available at GitHub.
In this subsection we describe its implementation and the challenges we had to tackle.

The algorithm for testing BFB conditions of a given scalar potential $V$ includes the following steps:
\begin{enumerate}
    \item Rewrite the potential $V$ in terms of real scalar fields $\vec{x} \in \mathbb{R}^n$, 
    extract the tensor of quartic couplings $Q_{i j k l}$, and set up the polynomials 
    $f_i = Q_{i j k l} \cdot x_j x_k x_l - \lambda \cdot x_i^3$.
    \item Calculate $\textrm{Char}(Q, \lambda) = \textrm{Res}(f_1, f_2, \dots, f_n)$.
    \item Find all real roots $\lambda \in \mathbb{R}$ of $\textrm{Char}(Q, \lambda) = 0$.
    \item Check all non-positive roots $\lambda \le 0$ for non-trivial, real solutions $\vec{x} \in \mathbb{R}^n \setminus \{0\}$ to the equations $f_1 = f_2 = \dots = f_n = 0$.
    \item The potential $V$ is bounded from below if and only if there are no such real solutions for the non-positive roots.
\end{enumerate}

In step 1, it is important to make sure that all real fields $x_i$ can span the entire real space and not just a subset of it.
If this condition is not met, the algorithm may only yield sufficient but not necessary constraints on the scalar potential parameters, 
simply because positive definiteness of $Q$ in the entire space may be too restrictive. 
It is this requirement that impedes its application in the space of gauge-invariant bilinears in multi-Higgs-doublet models.

Step 2 is the key step of the algorithm and it is more complicated.
The usual computer-algebra packages such as Mathematica and Maple have implementations for the calculation of resultants for two polynomials in at most two variables.
They do not have a general implementation for the calculation of multivariate resultants.
One way to proceed would be to implement the resultant algorithm presented in Section~\ref{sec:bounded:resultantalgo} 
within Mathematica or Maple, relying on their support for polynomial division algorithms such as finding Gröbner bases etc.
An alternative procedure is to use a more specialized computer algebra system such as Macaulay2 \citep{M2}, 
which is designed for problems in algebraic geometry.
It allows for the symbolic manipulation of polynomials and the calculation within quotient rings and ideals over the field of integers or rational numbers.
The implementation of multivariate resultants is provided as a package called Resultants \citep{ResultantsSource}.
The currently tested version of BFB \citep{BFB} uses this package.

At step 3, for analytic Higgs potential parameters, it is not clear if it is in general possible to decide whether a root is real or not.
Hence, most of the time this has to be decided after numeric values have been chosen.

Similarly to the calculation of resultants, performing step 4 can be rather involved.
It reduces to a proof of existence of real solutions for a given set of polynomial equations.
In the univariate case this problem can be tackled by the Sturm sequence.
For the more interesting multivariate case, the decision problem of real solutions has been solved 
by the Tarski-Seidenberg theorem \citep{MathOverflow}.
The implementation of BFB uses Mathematica's function called FindInstance to construct a real solution if possible.

In practice there are two different scenarios for which one would apply this algorithm.
Firstly, to have a numerical check of boundedness for a given point in the parameters space of the Higgs potential.
The Higgs potential will have numeric coefficients from the beginning.

Secondly, one would want to derive analytic constraints that can be later evaluated numerically.
The algorithm in step 2 can in principle produce the characteristic polynomial $\textrm{Char}(Q, \lambda)$ in analytic form for any model.
However, because calculating resultants is \mbox{NP-hard}~\citep{NPhardResultant}, this step can be very challenging.
We saw that calculating the resultant in analytic form within the IDM, which is discussed below, easily exceeds the time scale of several weeks with the current implementation of BFB.
We are confident that this implementation is not the most optimal one, and we hope that more efficient algorithms can be applied.

Next, even if the characteristic polynomial $\textrm{Char}(Q, \lambda)$ is known in analytic form,
its degree can easily grow far above four, which may preclude expressing its roots in an analytic way. 
Thus, at this stage, one would need to resort to numerical methods and explore the parameter space with numerical scans.
Fortunately, numerically solving a polynomial equation in a single real variable can be done in relatively short time.
We found that, for the IDM case, numerical calculations in step 3 and 4 take at most a few seconds.

\subsection{Inert Doublet Model}\label{sec:bounded:proof}

The quartic part of the Higgs potential of the Inert Doublet Model, 
which makes use of two Higgs electroweak doublets $\phi_1, \phi_2 \in \mathbb{C}^2$, 
is given by Eq.~(\ref{IDM-potential}).
The analytic BFB conditions were first derived in \citep{IDMBFBConditions} and are given in (\ref{IDM-BFB}).

We treated this problem with the BFB package \citep{BFB}.
The scan of the parameter space was done numerically. This means that we did not attempt to derive the analytical expression of the characteristic polynomial but, for each point in the scan, 
the Higgs potential parameters were assigned numerical values before running the algorithm.
To reduce the complexity of the problem, the $\SU(2)$ symmetry of the Higgs potential was exploited.
A given potential value $V$ at a certain point $\vec{x} \in \mathbb{R}^8$ of variables 
can be equally expressed by a different point $\vec{x}~' \in \mathbb{R}^8$ 
if the two points are connected through an $\SU(2)$ transformation of the two Higgs doublets.
Hence by an appropriate choice of transformation, one can make three of the eight variables vanish.
This eliminates flat directions of the potential and corresponds to the calculation of constraints in unitary gauge.

\begin{figure}[t]
    \captionsetup[subfigure]{%
        position=top,
        labelfont=bf,
        textfont=normalfont,
        singlelinecheck=off,
        justification=raggedright
    }%
    \centering
    \begin{subfigure}{0.45\textwidth}
        \centering
        \caption{}
        \includegraphics[width=0.95\textwidth,keepaspectratio]{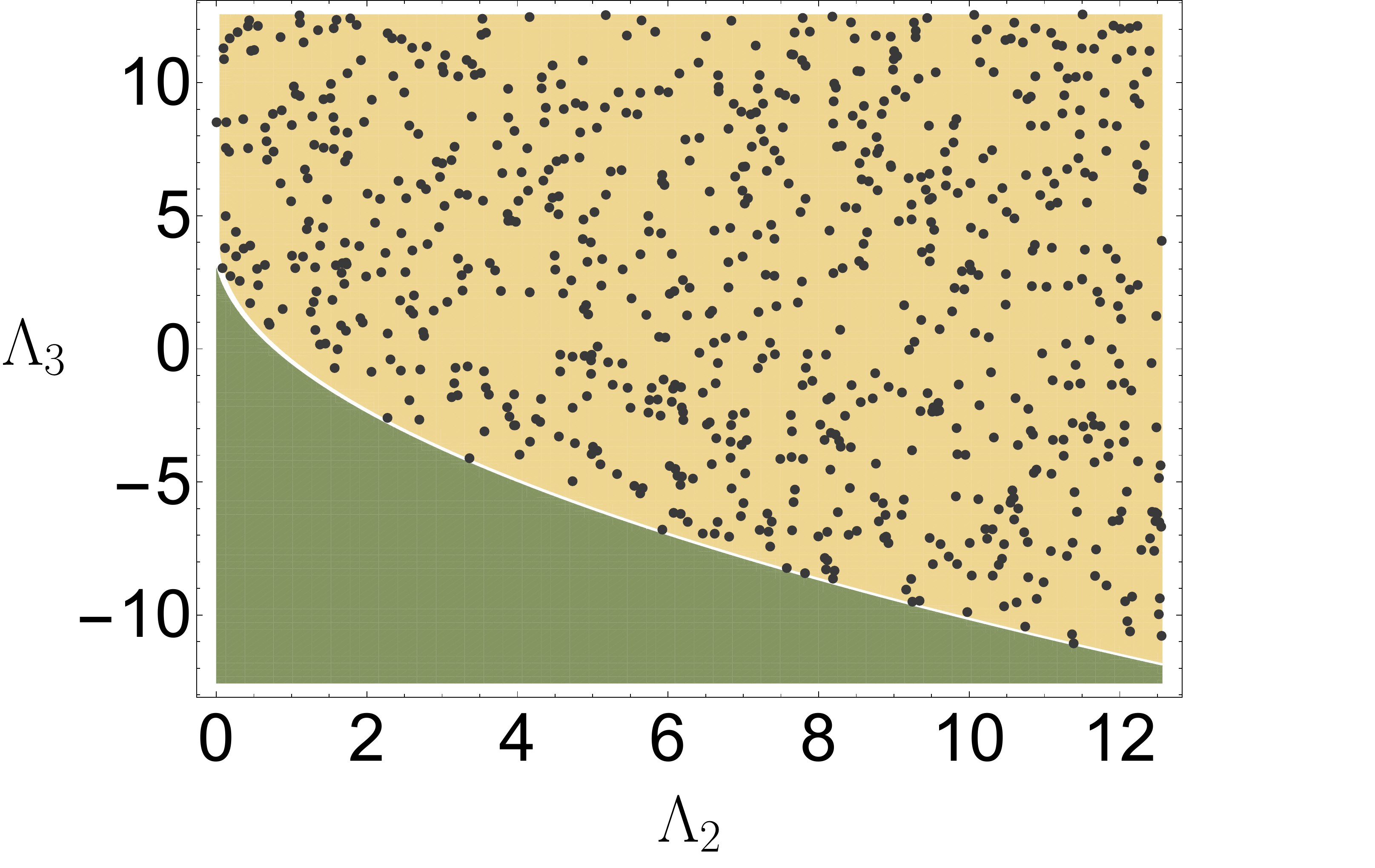}
    \end{subfigure}
    \begin{subfigure}{0.45\textwidth}
        \centering
        \caption{}
        \includegraphics[width=0.95\textwidth,keepaspectratio]{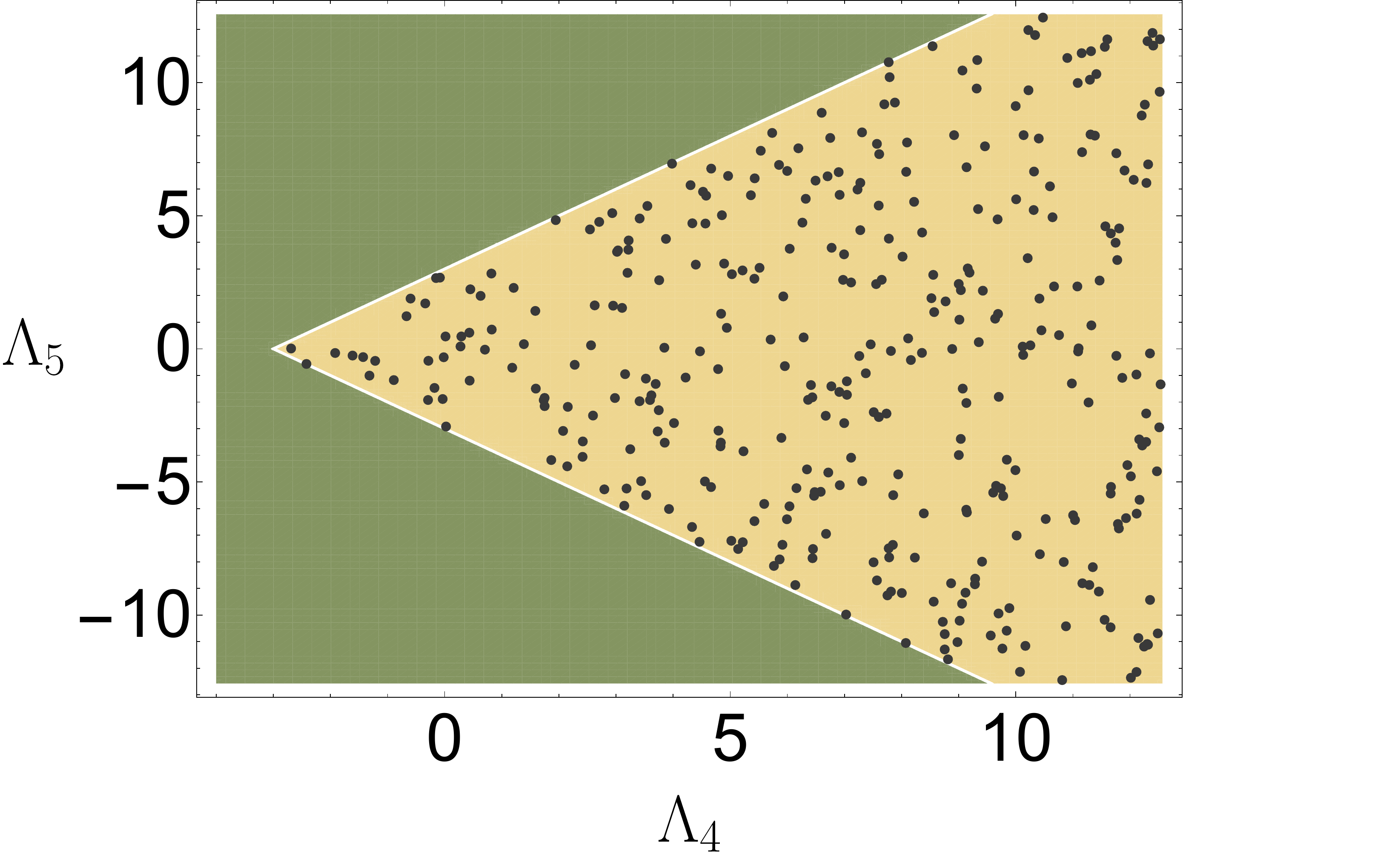}
    \end{subfigure}
    
    \caption{Exclusion plot for two parameter planes: (i) with $\Lambda_1 = \Lambda_5 = 5$, $\Lambda_4 = 1$ and (ii) with $\Lambda_1 = \Lambda_2 = \Lambda_3 = 1$. The green region is excluded by analytic constraints, the yellow region is allowed. Black points are allowed according to a parameter scan with BFB~\citep{BFB}.}
    \label{fig:IDMbounded}
\end{figure}

In Fig.~\ref{fig:IDMbounded} we show the exclusion plots in a selection of two parameter planes;
additional plots can be found in \cite{MarcelThesis}.
The green region is excluded by the analytic constraints of Eq.~\eqref{IDM-BFB}.
The yellow region is allowed.
Black dots are those points from the numerical scan which were approved by the package BFB. 
They perfectly agree with the analytical conditions.

It is worth mentioning that Macaulay2 \citep{M2} allows one to calculate the resultant not only over a field but also over the ring of integers.
This is on average faster because the intermediate polynomial division steps require the division of the coefficients.
Within the ring of integers this is effectively done by a modulo operation which is much faster than an actual division.
The computation time varied from 3 hours to 8 hours per parameter space point.
Almost the whole time was spent for the calculation of the resultant.
We also observed a strong dependence of the calculation time on the complexity of the input parameters:
simpler coefficients such as $1/10$ would result in a faster calculation than coefficients like $743/999$.

According to (\ref{counting-eigenvalues}), for the five-variable version of the IDM, 
the degree of the characteristic polynomial is equal to $N_Q = 5 \times 3^4 = 405$.
Hence the initial parameters will approximately be raised to this total power 
making the resulting numerator and denominator a huge number which cannot be stored in CPU registers.
For instance, with $\Lambda_1 = \Lambda_2 = \Lambda_3 = 1$, $\Lambda_4 = 9.01587 $ and $\Lambda_5 = - 10.2132$ the largest coefficient of the characteristic polynomial is
\begin{equation}
    \approx 3.452 \cdot 10^{1137}\,.
\end{equation}
Thus, special libraries for integer manipulation, which emulate the CPU's arithmetic logic unit, have to be used.
The runtime of the remaining algorithm, after the calculation of the characteristic polynomial, is negligible.
For the above parameter point, calculating and testing all H-eigenvalues takes no longer than $3$ seconds.

%% file: sec/discussion.tex
\subsection{The Present Situation}

Checking that scalar potentials are BFB is a notoriously difficult problem,
which impedes efficient exploration of many models with extended scalar sectors.
Analytical BFB conditions are known only in special and rather simple cases.
For example, in models with three Higgs doublets the BFB conditions remain unknown
beyond the few cases with large symmetry groups.

In this work we presented and developed a novel approach to establishing the BFB conditions of generic polynomial scalar potentials, which, to our best knowledge, was briefly mentioned only in \cite{Kannike:2016fmd} 
and was not pursued any further by the HEP community.
The method relies on certain unconventional mathematical methods such as the theory of resultants 
and the spectral theory of tensors. In this approach, the BFB conditions are equivalent to calculating
a well defined characteristic polynomial and checking that its real roots satisfy certain conditions.
We described an explicit algorithm of calculating the characteristic polynomial and illustrated 
it with two elementary cases, where all calculations can be done manually.
We also implemented the algorithm in a Mathematica package BFB \cite{BFB} which is publicly available at GitHub.
We validated its performance with the case of the Inert Doublet Model, for which the conditions
are known analytically, and we found perfect agreement. 

Unfortunately, we have not yet produced ready-to-use analytical results for other, more complicated cases, 
where the BFB conditions are at present unknown.
This is in part due to the intrinsic complexity of the problem:
it is NP-hard and the computation time grows exponentially with the input information.
However, we also believe that our current implementation is not the most optimal one,
and we hope that it can be dramatically improved in the future.
Since the approach is novel, we call for a community effort in optimizing this approach.

\subsection{Directions for Future Work}

The algorithm presented in this work is capable of constructing BFB constraints for any Higgs potential.
The bottleneck of runtime is the calculation of the characteristic polynomial.
We see four possible improvements that may increase the speed drastically.

First, the current implementation of BFB \citep{BFB} uses no parallelization even though there is great potential to do so.
This is mainly because the whole calculation of the resultant is outsourced to the computer algebra system Macaulay2 \citep{M2}.
There are two critical algorithms that may be subject to improvement: the calculation of the Gröbner bases and the calculation of the resultant.
Both of them are under steady investigation of the mathematical community.
For Gröbner bases, there are Faugére's algorithms F4 \citep{F4Algo} and F5 \citep{F5Algo} both of which are highly parallelizable.
Macaulay2 includes already four different algorithms for the calculation of the resultant.
The algorithm presented in Section \ref{sec:bounded:resultantalgo} from \citep[theorem 3.4]{UsingAlgebraicGeometry} is one of them.
Part of it is the calculation of the intermediate matrices $M_i$.
Currently, the elements are obtained in a linear way on one CPU only.
However, each row can be calculated independently.
For the IDM test of Section \ref{sec:bounded:proof}, $M_1$ already has $81$ rows, so here is a huge potential for parallelization.
Also, the calculation of the basis of the quotient ring is a simple scan through low degree polynomials and can be distributed over any number of cores.
Macaulay2 implements also the classic algorithm by Macaulay \citep{Macaulay1902}.
It is less space efficient but may be more time efficient when it comes to the calculation of resultants of polynomials with many variables.
Furthermore, Macaulay2 implements a variation of these two algorithms that makes use of polynomial interpolation (see for instance \citep{ResultantAlgoInterpolation2} and \citep{ResultantAlgoInterpolation}).

Second, as one can probably already conclude, not only the possibility of parallelization 
may speed up the process of resultant calculations, but also the choice of the respective algorithm.
There is a multitude of publications on this topic.
Depending on the specific form of the input polynomials there might exist much faster algorithms than the presented one.
For instance, Macaulay proposed a modified version of his algorithm that can be used if all polynomials share the same degree \citep{ResultantAlgoSameDegree}.
This is applicable to the current case of Higgs potential boundedness and should definitely be tested.
It is this approach that might bypass the NP-hardness \citep{NPhardResultant} of resultant calculations for Higgs potential boundedness.

Third, the scalar potentials we encounter are gauge invariant, and this implies a certain redundancy when writing them in terms of real fields.
For example, for the IDM test of Section \ref{sec:bounded:proof}, we used the $\SU(2)$ symmetry of the Higgs potential 
to reduce the number of variables from 8 to 5.
It is plausible that additional symmetries of other multi-Higgs models can be exploited in a similar way.
Furthermore, since the BFB check can be performed in any basis,
one may take advantage of the basis-change freedom to switch to a basis that is more convenient.
This may result in a further reduction of variables or parameters.
Another symmetry driven approach is the usage of E-eigenvalues \citep{QiReview}, which are, unlike H-eigenvalues, invariant under orthogonal transformations.
A short discussion of the implications with respect to Higgs potential boundedness can be found in \citep{MarcelThesis}.

Lastly, when performing the scans of the parameter space, one can use the ring of integers instead of a field of numbers for the polynomial coefficients.
As we saw with the IDM example in Section \ref{sec:bounded:proof}, this option changes the runtime.
Rational numbers $\mathbb{Q}$ might be the worst choice because they incorporate an inefficient division algorithm 
(finding greatest common divisors etc.) and have a bad scaling with powers (numerator and denominator can get very large).
Integers $\mathbb{Z}$ are more efficient when it comes to the used division operations (modulo operations) but still possess a bad scaling with powers.
Macaulay2 only allows for these two options.
The field of real numbers $\mathbb{R}$ may be an intermediate solution that trades accuracy for runtime.
The division algorithm is not as fast as for integers but calculations of powers are faster and more space efficient (floating point numbers store powers separately).
Currently there exists no implementation of resultant algorithms that work with both analytic parameters and real numbers.
There is a working framework called MARS \citep{MARS} that can handle the calculation of the resultant numerically.
It is possible to perform a scan over a bounded range of values for the eigenvalues $\lambda$ and test for the numerical vanishing of the resultant.
This is numerically unstable though, since the resultant is in general a high-degree polynomial in $\lambda$ and accuracy will play an important role here.
Nevertheless, this is a feasible approach.

The long term goal is to have an algorithm that can produce the analytic form of the characteristic polynomial for various Higgs potentials.
It is true that computing this polynomial in a specific model, for example, in 3HDM, even after parallelization and optimizaiton may require much computer time.
However, once the characteristic polynomial is calculated in its full analytic form,
it can be published and distributed, and it can be readily used for all subsequent checks of BFB conditions in this model.
Such ``mining of characteristic polynomials'' is definitely worthy of extra efforts.

%% file: acknowledgments.tex
We are grateful to Kristjan Kannike for his valuable comments.
We also want to thank Sven Caspart for the insightful discussions about algebraic geometry.
I.P.I. was supported by the Portuguese 
\textit{Fun\-da\-\c{c}\~{a}o para a Ci\^{e}ncia e a Tecnologia} (FCT) 
through the Investigator contract 
IF/00989/2014/CP1214/CT0004 under the IF2014 Program 
and in part by contracts 
UID/FIS/00777/2013 and CERN/FIS-NUC/0010/2015, 
which are partially funded through POCTI, COMPETE, QREN, and the European Union.
M.K. and M.M. acknowledge financial support from the DFG project “Precision Calculations in the
Higgs Sector --- Paving the Way to the New Physics Landscape” (ID: MU 3138/1-1).

%% file: app/resultants.tex
In this section, we remind the reader of basic notions on the ring of polynomials,
their division, and the theory of resultants, the objects which indicate when a system of polynomial
equations has non-trivial solutions. To keep the material easily readable, 
we will expose it in plain language and reduce the formal notation to the bare minimum.

\subsection{Polynomial Rings}
A polynomial in one variable $x$ is an expression of the form
\begin{equation}
    g := a_0 + a_1 \cdot x + a_2 \cdot x^2 + \dots + a_m\cdot x^m\,,
\end{equation}
with the coefficients $a_i$ belonging to some field $\mathbb{F}$, e.g. $\mathbb{Q}$, $\mathbb{R}$, or $\mathbb{C}$.
The non-negative integer $m$ is called the degree of the polynomial.
The collection of all polynomials of all degrees forms an algebraic structure called the polynomial ring $\mathbb{F}[x]$.
A ring has richer structure than a vector space, because its elements, in addition to summation and multiplication
by another number from the field $\mathbb{F}$,
can also be multiplied among themselves, with the result staying inside the ring.
However, unlike fields, polynomial rings may lack multiplicative inverses.

Similarly, one can define a polynomial ring in several variables $x_1, x_2, \dots, x_n$ over the same field;
it is denoted as $\mathbb{F}[x_1, x_2, \dots, x_n]$. 
The polynomial $f \in \mathbb{F}[x_1, x_2, \dots, x_n]$ is called multivariate, 
while $g \in \mathbb{F}[x]$ is univariate.
A monomial in $\mathbb{F}[x_1, x_2, \dots, x_n]$ is an expression of the form $x_1^{d_1} x_2^{d_2} \dots x_n^{d_n}$.
Its (total) degree is the sum of all individual powers: $d = \sum_i d_i$.
Clearly, any polynomial is a linear combination of monomials.
The degree of a multivariate  polynomial is the highest degree among all of its monomials. 

\subsection{Ideals and Polynomial Division}

Just as for groups or vector spaces, rings can have subrings, which are subsets closed under all of its operations.
However, rings also contain another important substructures called {\em ideals}. 
A polynomial ideal is a subring $I \subseteq \mathbb{F}[x]$ such that $I$ is closed under multiplication by the {\em whole} ring:
if $i \in I$ and $r \in \mathbb{F}[x]$, then $r \cdot i \in I$.

Ideals are closely related to polynomial division. Consider first univariate polynomials.
The following theorem holds: for all $f, g \in \mathbb{F}[x]$, there are unique $q, r \in \mathbb{F}[x]$ such that
\begin{equation}
    f = q \cdot g + r\,,
\end{equation}
with either $r = 0$ or $\deg(r) < \deg(g)$.
One calls $q$ the quotient and $r$ the remainder of the (Euclidean) polynomial division of $f$ by $g$. 
If $r=0$, then $f$ is divisible by $g$. The set of all $f$ that are divisible by $g$ forms an ideal $I$,
which is denoted as $I := \langle g \rangle$. By Hilbert's basis theorem, every univariate ideal has this form,
see, for example, \citep[p. 4]{UsingAlgebraicGeometry}.

For a multivariate polynomial ring $R := \mathbb{F}[x_1, x_2, \dots, x_n]$, every ideal is also of the form 
\begin{equation}
    I = \langle g_1, g_2, \dots, g_m \rangle = \left\{ q_1 \cdot g_1 + q_2 \cdot g_2 + \dots + q_m \cdot g_m ~ | ~ q_i \in R \right\}
\end{equation}
for some $m \in \mathbb{N}$. The polynomials $g_i$ are called generators of $I$.
Polynomial ideals are always finitely generated.
However, the relation of ideals to polynomial division, that is, representing $f$ as
\begin{equation}
    f = q_1 \cdot g_1 + q_2 \cdot g_2 + \dots + q_m \cdot g_m + r\,,\label{multivariate-division}
\end{equation}
with some remainder $r$ becomes more subtle.

Consider, for example, $\mathbb{F}[x_1, x_2]$ and try to divide $f = x_1^2 x_2 + x_1 x_2^2 + x_2^2$
by $g_1 = x_1 x_2 - 1$ and $g_2 = x_2^2 - 1$.
Then, the decomposition (\ref{multivariate-division}) is not unique:
\begin{equation}
\label{equ:bounded:divisionexample}
\begin{alignedat}{3}
    f & = (x_1 + x_2) \cdot g_1 && + 1 \cdot g_2 && + (x_1 + x_2 + 1) \\
    f & = x_1 \cdot g_1 && + (x_1 + 1) \cdot g_2 && + (2 x_1 + 1)\,.
\end{alignedat}
\end{equation}
The uniqueness of the $q_i$ and $r$ is lost compared to the univariate case.
To understand why this happened one has to look at the explicit algorithm used to derive these results.
In the first example of Eq.~\eqref{equ:bounded:divisionexample}, the term $(x _1+ x_2) \cdot g_1$ cancels 
both of the terms of highest degree in $f$, while in the second example $x _1\cdot g_1$ only cancels the first one.
For univariate polynomials there is no ambiguity in deciding what the leading order term is.

To restore the uniqueness for the multivariate case, one first has to introduce a monomial ordering, 
which would uniquely identify a leading term for every polynomial.
Several options are possible. For example, the lexicographic order starts by ordering the variables themselves,
$x_1 > x_2 > \dots > x_n$, and then demands that 
\begin{equation}
\label{equ:bounded:monomialorder}
    x_1^{d_1} x_2^{d_2} \dots x_n^{d_n} > x_1^{e_1} x_2^{e_2} \dots x_n^{e_n}
\end{equation}
if in the difference $(d_1,\dots,d_n) - (e_1,\dots,e_n)$ the left-most non-zero entry is positive.
This is analogous to the ordering of words in dictionaries.

Still, choosing the lexicographic ordering does not completely restore the uniqueness in the above example.
However, instead of thinking of Eq.~(\ref{multivariate-division}) as a division of $f$ by a given set of polynomials $g_i$,
one can think of it as a division by the ideal $\langle g_1, \dots, g_m\rangle$.
In this way, one can define an equivalent set of polynomials, which span the same ideal and for which uniqueness is restored.
That is, for every set of $g_i$, there exists a set of $\tilde{g}_i$ called a Gröbner basis such that
\begin{equation}
    I = \langle g_1, g_2, \dots, g_m \rangle = \langle \tilde{g}_1, \tilde{g}_2, \dots, \tilde{g}_{\tilde{m}} \rangle
\end{equation}
and it holds that for any $f \in I$ there exists a $\tilde{g}_i$ such that the leading term of $f$ is divisible by the leading term of $\tilde{g}_i$.
In a sense, a Gröbner basis is the smallest generating set for $I$;
it is a convenient choice to make division unique.
There exists an algorithm due to Buchberger (see for example \citep[p. 15]{UsingAlgebraicGeometry}), which allows one to find a Gröbner basis algorithmically.
Most computer algebra systems like Mathematica (function call: GroebnerBasis) and Maple (package: Groebner, function call: Basis) implement this or similar algorithms.

In the above example (\ref{equ:bounded:divisionexample}) with the lexicographic ordering, the Gröbner basis for $\langle g_1, g_2\rangle$ 
is given by $\tilde g_1 = x_1 - x_2$, $\tilde g_2 = g_2 = x_2^2 -1$.
The ideals spanned by both pairs are equal because $g_1 = x_2 \tilde g_1 + \tilde g_2$
and, conversely, $\tilde g_1 = x_2 g_1 - x_1 g_2$.
The division of $f$ by $\langle g_1, g_2\rangle$ can be performed as a division by polynomials
$\tilde g_i$ using the lexicographic ordering, resulting in
\be
f = (x_1x_2+2x_2^2) \tilde g_1 + (2 x_2 + 1) \tilde g_2 + 2x_2+1\,.
\ee
A different ordering may lead to a different Gröbner basis and a different remainder.

\subsection{Quotient Ring as a Vector Space}\label{app:subsection-quotient}

Consider a polynomial ring $R = \mathbb{F}[x_1, x_2, \dots, x_n]$ and an ideal $I = \langle g_1, g_2, \dots, g_m \rangle$,
where $g_i$ already indicate its Gröbner basis.
Every polynomial $f \in R$ can be uniquely divided by the ideal $I$ producing a remainder $r$, see Eq.~(\ref{multivariate-division}).
Different polynomials $f_1$ and $f_2$ can produce the same remainder $r$, if $f_1 - f_2 \in I$.
Therefore, one can consider the remainder $r$ as the smallest representative of an equivalence class of remainders, denoted by $[r]$,
which represents all polynomials $f \in R$ such that their division by the ideal $I$ gives $r$.

The collection of all such equivalence classes also forms a ring called the quotient ring 
\be
Q = \mathbb{F}[x_1, x_2, \dots, x_n] / \langle g_1, g_2, \dots, g_m \rangle\,.
\ee
Calculations in it are the same as calculations in $\mathbb{F}[x_1, x_2, \dots, x_n]$ modulo the ideal $I$:
$[r_1] + [r_2] = [r_1 + r_2]$, $[r_1] \cdot [r_2] = [r_1 \cdot r_2]$.
Practically, one takes two representatives $f_1 \in [r_1]$ and $f_2 \in [r_2]$, 
performs the calculations, and divides the outcome by $I$.

In certain situations, one can also view the quotient ring $Q$ as a $\mathbb{C}$ vector space spanned by a {\em finite} monomial basis.
This is the case for the quotient ring used in \eqref{equ:bounded:resultantmap}, i.e. for calculating resultants.
Let $D := \dim Q$ and consider now any polynomial $f \in \mathbb{F}[x_1, x_2, \dots, x_n]$, which is a representative element of the equivalence class $[f]$.
By multiplying elements $[r] \in Q$ by $[f]$, one obtains other elements $[r \cdot f]$, which also belong to the same vector space $Q$.
Thus, $f$ induces a linear map $M_f$ in the vector space $Q$.
If a basis $\{[r_a]\}$ is chosen in the vector space $Q$, one can describe this map with a matrix $M$ acting on the basis vectors according to
\be
[r_a] \to [r_a \cdot f] = \sum_{b=1}^{D} ~ [r_b]\cdot M_{b a}\,.
\ee

\subsection{Working with Resultants}
\label{sec:bounded:resultantexample}

To get some practice with quotient space calculations, which are needed for the algorithm presented in Section~\ref{sec:bounded:resultantalgo}, let us consider the following system of three homogeneous polynomials
\begin{equation}
\label{equ:bounded:fexampleb}
\begin{split}
    f_1 & = x^3 - x y z + y^2 z \\
    f_2 & = x^2 + y z \\
    f_3 & = y^2 + z^2
\end{split}
\end{equation}
We construct two other sets of polynomials:
\begin{equation}
\label{exampleb-fF}
\begin{alignedat}{2}
    & \bar{f}_1 = f_1(0,y,z) = y^2z\,,\qquad    && F_1 = f_1(1,y,z) = y^2 z - y z + 1\,, \\
    & \bar{f}_2 = f_2(0,y,z) = yz\,,\qquad      && F_2 = f_2(1,y,z) = y z + 1\,, \\
    & \bar{f}_3 = f_3(0,y,z) = y^2 +z^2\,,\quad && F_3 = f_3(1,y,z) = y^2 + z^2\,.
\end{alignedat}
\end{equation}
We then build the ideal $\langle F_2, F_3 \rangle$ and,
adopting the lexicographic ordering, find its Gröbner basis:
\begin{equation}
    G_2 = z^4 + 1\,,\qquad
    G_3 = y - z^3\,.
\end{equation}
The quotient ring and vector space $\mathbb{C}[y,z] / \langle F_2, F_3 \rangle$ has dimension $D = {\deg(f_2) \cdot \deg(f_3)} = 4$. 
Table \ref{tab:bounded:remainders} shows a list of remainders $r$ for a division of low degree monomials.
\begin{table}
\centering
\caption{Remainders $r$ of a polynomial division of low degree monomials by the Gröbner basis $G_2, G_3$.}
\begin{tabular}{l c c c c c c c c c c}
    \hline
    monomial  & $1$ & $y$   & $z$ & $y^2$   & $y z$ & $z^2$ & $y^3$ & $y^2 z$ & $y z^2$ & $z^3$ \\
    remainder $r =$ & $1$ & $z^3$ & $z$ & $- z^2$ & $-1$  & $z^2$ & $z$   & $- z^3$ & $- z$   & $z^3$ \\
    \hline
\end{tabular}
\label{tab:bounded:remainders}
\end{table}
It allows us to select the four basis vectors $[r_1] = [1]$, $[r_2] = [z]$, $[r_3] = [z^2]$ and $[r_4] = [z^3]$.
Now consider the products $r_a \cdot F_1$ and perform their polynomial division by $G_2, G_3$: 
\begin{equation}
\begin{alignedat}{3}
    r_1 \cdot F_1 & = (y - 1) \cdot G_2 && + (y z - z - 1) \cdot G_3 && + \ (2 - z^3) \\
    r_2 \cdot F_1 & = (y z - z - 1) \cdot G_2 && + (y z^2 - z^2 - z) \cdot G_3 && +\ (1 + 2 \cdot z) \\
    r_3 \cdot F_1 & = (y z^2 - z^2 - z) \cdot G_2 && + (y z^3 - z^3 - z^2) \cdot G_3 && +\ (z + 2 \cdot z^2) \\
    r_4 \cdot F_1 & = (y^2 - z^2 - y) \cdot G_2 && + (- z^3 - y + 1) \cdot G_3 && +\ (z^2 + 2 \cdot z^3)
\end{alignedat}
\end{equation}
and expand the remainders in the same basis vectors $[r_a]$:
\begin{equation}
\begin{alignedat}{5}
    [r_1 \cdot F_1] & = [2 - z^3] && = 2 \cdot [r_1] && && && - 1 \cdot [r_4] \\
    [r_2 \cdot F_1] & = [1 + 2 \cdot z] && = 1 \cdot [r_1] && + 2 \cdot [r_2] && && \\
    [r_3 \cdot F_1] & = [z + 2 \cdot z^2] && = && + 1 \cdot [r_2] && + 2 \cdot [r_3] && \\
    [r_4 \cdot F_1] & = [z^2 + 2 \cdot z^3] && = && && + 1 \cdot [r_3] && + 2 \cdot [r_4]\,.
\end{alignedat}
\end{equation}
Thus, $F_1$ acts in this vector space with the matrix $M_1$ given by
\begin{equation}
    M_1 = \begin{pmatrix}
    2  & 1 & 0 & 0 \\
    0  & 2 & 1 & 0 \\
    0  & 0 & 2 & 1 \\
    -1 & 0 & 0 & 2
    \end{pmatrix}
\end{equation}
with determinant $\det(M_1) = 17$.
Therefore, the first step of the algorithm of Section~\ref{sec:bounded:resultantalgo} gives
\be
\Res(f_1,f_2,f_3) = \left( ~ \Res(\bar{f}_2,\bar{f_3}) ~ \right)^3 \cdot 17\,.
\ee
The second step starts with $\bar{f}_2, \bar{f}_3$ in Eq.~(\ref{exampleb-fF}),
from which we construct
\begin{equation}
\begin{alignedat}{2}
    & \tilde{f}_2 = 0\,,\qquad  && \tilde{F}_2 = z\,, \\
    & \tilde{f}_3 = z^2\,,\quad && \tilde{F}_3 = 1 + z^2\,.
\end{alignedat}
\end{equation}
The quotient ring $\mathbb{C}[z] / \langle \tilde{F}_3 \rangle$ viewed as a vector space has dimension $\deg(\bar{f}_3) = 2$,
and its basis vectors are $[1]$ and $[z]$.
In this space, $\tilde{F}_2$ acts as a linear map with the following matrix:
\begin{equation}
    M_2 = \begin{pmatrix}
     0 & -1 \\
    1 & 0
    \end{pmatrix}\,.
\end{equation}
We have $\det(M_2) = 1$.
Finally, using that $\Res(\tilde{f}_3) = 1$, we obtain
$\Res(\bar{f}_2,\bar{f_3}) = 1^2 \cdot 1 = 1$, so that the overall resultant 
of the system~(\ref{equ:bounded:fexampleb}) is equal to $\Res(f_1,f_2,f_3) = 17$.
Since it is non-zero, the system of equations $f_i = 0$ does not possess non-trivial solutions.